\documentclass[twocolumn,showpacs,preprintnumbers,amsmath,amssymb,nofootinbib,scrartcl]{revtex4}
\input psfig.sty
\def \be {\begin{equation}}
\def \ee {\end{equation}}
\def \bea {\begin{eqnarray}}
\def \eea {\end{eqnarray}}

\usepackage{graphicx}% Include figure files
\usepackage{dcolumn}% Align table columns on decimal point
\usepackage{bm}% bold math

\begin{document}

\title{Thermodynamics and Dark Energy}

\author{R. Silva$^{1}$}
\email{raimundosilva@dfte.ufrn.br}

\author{R. S. Gon\c{c}alves$^{2}$}
\email{rsousa@on.br}

\author{J. S. Alcaniz$^{2}$}
\email{alcaniz@on.br}

\author{H. H. B. Silva$^{1}$}
\email{heydson@dfte.ufrn.br}

\affiliation{$^{1}$Universidade Federal do Rio Grande do Norte, Departamento de F\'{\i}sica, Natal - RN, 59072-970, Brasil}

\affiliation{$^{2}$Observat\'orio Nacional, Rua General Jos\'e Cristino 77, 20921-400 Rio de Janeiro - RJ, Brasil}


\date{\today}

\begin{abstract}
A significant observational effort has been directed to unveil the nature of  the so-called dark energy. However, given the large number of theoretical possibilities, it is possible that such a task cannot be performed on the basis only of the observational data. In this article we discuss some thermodynamic properties of this energy component assuming a general time-dependent equation-of-state parameter $\omega(z)=\omega_0 + \omega_a f(z)$, where $\omega_0$ and $\omega_a$ are constants and $f(z)$ may assume different forms. We show that very restrictive bounds can be placed on the $w_0$ - $w_a$  space when current observational data are combined with the thermodynamic constraints derived.
\end{abstract}

\maketitle

\section{Introduction}

The observational evidence for an acceleration of the expansion of the Universe is now overwhelming, although the precise cause of this phenomenon is still unknown (see, e.g., \cite{review} for recent reviews). In this concern, and besides the need for more accurate estimates of cosmological parameters, the current state of affairs also brings to light some other important aspects regarding the physics of the mechanism behind cosmic acceleration. Certainly, one of these aspects concerns the thermodynamical behavior of a dark energy-dominated universe, and questions such as ``what is the thermodynamic behavior of the dark energy in an  expanding universe?'' or, more precisely, ``what is its temperature evolution law?'' must be answered in the context of this new conceptual set up. Another interesting aspect in this discussion is whether thermodynamics in the accelerating universe can place constraints on the time evolution of the dark energy and can also reveal some physical properties of this energy component.

The aim of this paper is twofold. First, to derive physical constraints on the dark energy from the second law of thermodynamics and to deduce the temperature evolution law for a dark component with a general equation-of-state (EoS) parameter $w(a)$; Second, to perform a joint statistical analysis involving current observational data together with the thermodynamic bounds on $w(a)$. To do that, we assume the following generalized formula for the time evolution of $w(a)$~\cite{geral}
\begin{eqnarray} \label{pb}
\label{pbeta}
w(a) \equiv {p_x \over \rho_x} & = & w_0+w_{a}\frac{1 - a^{\beta}}{\beta}
\nonumber \\
& = &w_0+w_{a}\frac{1 - (1+z)^{-\beta}}{\beta} \;,
\end{eqnarray}
which recovers some well-known EoS parameterizations in the following limits:
\begin{eqnarray}
\label{ps}
 w(z) = \left\{
 \begin{tabular}{l}
 $w_0 + w_{a}{(1 - a) \over a}$ \quad \quad \hspace{0.05cm} (P1) \quad $\beta \rightarrow -1$ \quad\hspace{0.13cm}\cite{8}\\
\\
$w_0-w_{a}\ln{a}$ \quad  \quad \quad (P2) \quad $\beta \rightarrow 0$ \quad\quad\hspace{0.10cm}\cite{14}\\
\\
$w_0+w_{a}{(1-a)}$ \quad  \hspace{0.15cm} (P3) \quad $\beta \rightarrow +1$ \quad\hspace{0.14cm}\cite{15}\\
\end{tabular}
\right.
\nonumber
\end{eqnarray}
where $p_x$ and $\rho_x$ stand for the dark energy pressure and energy density, respectively (see also \cite{para} for other EoS parameterizations). The analyses are performed using one the most recent type Ia supernovae (SNe Ia) observations, the   nearby + SDSS + ESSENCE + SNLS + Hubble Space Telescope (HST) set of 288 SNe Ia discussed in Ref.~\cite{sdss} (which we refer to as SDSS compilation). We consider two sub-samples of this latter compilation that use SALT2~\cite{salt2} and MLCS2k2~\cite{mlcs2k2} SN Ia light-curve fitting method. Along with the SNe Ia data, and  to help break the degeneracy between the dark energy parameters we also use the baryonic acoustic oscillation (BAO) peak at $z_{\rm{BAO}} = 0.35$~\cite{bao} and the current estimate of the CMB shift parameter ${\cal{R}} = 1.71 \pm 0.019$~\cite{wmap}. We work in units where c = 1. Throughout this paper a subscript 0 stands for present-day quantities and a dot denotes time derivative.

\section{Thermodynamic Analysis}

Let us first consider a homogeneous, isotropic, spatially flat cosmologies described by the Friedmann-Robertson-Walker (FRW) flat line element, $ds^2=dt^2-a^2(t)(dx^2+dy^2+dz^2)$, where $a(t) = 1/(1+z)$ is the cosmological scalar factor. The matter content is assumed to be composed of baryons, cold dark matter and a dark energy component.

In such a background, the thermodynamic states of a relativistic fluid are characterized by an energy momentum tensor (perfect-type fluid)
\begin{subequations}
\begin{equation}
 T^{\alpha \beta}=\rho_x u^{\alpha} u^{\beta} - p_x h^{\alpha \beta}\;,
\end{equation}
a particle current,
\begin{equation}
N^{\alpha}=nu^{\alpha}\;,
\end{equation}
and an entropy current,
\begin{equation}
S^{\alpha}=n\sigma u^{\alpha}\;,
\end{equation}
\end{subequations}
%$T^{\alpha \beta}=\rho_x u^{\alpha} u^{\beta} - p_x h^{\alpha \beta}$, a particle current $N^{\alpha}=nu^{\alpha}$ and an entropy current $S^{\alpha}=n\sigma u^{\alpha}$,
where $h^{\alpha \beta}:=g^{\alpha \beta}-u^{\alpha} u^{\beta}$ is  the usual projector onto the local rest space of $u^\alpha$ and $n$ and $\sigma$ are the particle number density and the specific entropy (per particle), respectively~\cite{landau}. The conservation laws for energy  and particle number densities read
 \begin{subequations}
\begin{equation}\label{enmon}
u_\beta T^{\alpha \beta};_{\alpha}= {\dot \rho_x} + (\rho_x + p_x)\theta  = 0\;,
\end{equation}
 \begin{equation} \label{nalpha}
 N^{\alpha};_{\alpha}= \dot n + n\theta = 0\;,
\end{equation}
\end{subequations}
where semi-colons mean covariant derivative, $\theta = 3{\dot a} /a$ is the scalar of expansion and the quantities $p_x$, $\rho_x$, $n$ and $\sigma$ are related to the temperature $T$ trough the Gibbs law:
%\begin{equation} \label{eq:GIBBS1}
$nTd\sigma = d\rho_x - {{\rho_x + p_x} \over n}dn$.
%\end{equation}
From the energy conservation equation above, it follows that the energy density for a general $w(a)$ component can be written as
\begin{equation}
 \rho_{x } \propto \exp\left[{-3\int{\frac{1 + w(a)}{a}}da}\right]\;.
\end{equation}

Following standard lines (see, e.g., \cite{weinberg,silva}), it is  possible to show that the temperature evolution law is given by
\begin{equation} \label{eq:EVOLT}
{\dot T \over T} = \biggl({\partial p_0 \over \partial \rho_x}\biggr)_{n} {\dot n \over n} + \biggl({\partial \Pi \over \partial \rho_x}\biggr)_{n} {\dot n \over n} \;,
\end{equation}
where we have split the dark energy pressure as
\begin{equation}
p_x \equiv  p_0 + \Pi = w_0\rho_x +  w_{a}f(a)\rho_x\;,
\end{equation}
where  $f(a) = (1 - a^\beta)/\beta$. By combining the above equations, we also find  that
\begin{equation} \label{temp}
T \propto  \exp\left[{-3\int{\frac{ w(a)}{a}}da}\right]\;,
\end{equation}
where we have used that $n \propto a^{-3}$, as given by the conservation of particle number density [Eq. (\ref{nalpha})].
%\begin{equation}
%N^{\alpha};_{\alpha}= \dot n + 3n{{\dot a} \over a} = 0\;.
%\end{equation}
For parameterization P$_\beta$, shown in Eq. (\ref{pbeta}), Eq. (\ref{temp}) can be rewritten as
\begin{equation}
\label{Tpbeta}
T^x=  T^{x}_0a^{-3(w_0+{w_{a} \over \beta})}\exp\Big[-\frac{3w_{a}}{\beta}\big(\frac{1 - a^{\beta}}{\beta}\big) \Big]\;,
\end{equation}
which reduces to the generalized Stefan-Boltzmann law for $w_a = 0$~\cite{alcaniz} (see also~\cite{janilo}). From the above expressions, we confirm the results of Ref.~\cite{alcaniz} (for a constant EoS parameter) and find that dark energy becomes hotter in the course of the cosmological expansion since the its EoS parameter must be a negative quantity. A possible physical explanation for this behavior is that thermodynamic work is being done on the system (see, e.g., Fig. 1 of Ref.~\cite{alcaniz}). In particular, for the vacuum state ($w = -1$) we obtain $T \propto a^3$. To illustrate this behavior, Fig. 1 shows the dark energy temperature as a function of the scale factor for  $\beta = 0$ (P2) and $\beta = 1$ (P3)\footnote{As is well known, P1 has the drawback of becoming rather unphysical for $w_0 > 0$, with the dark energy density $\rho_x$ blowing up as $e^{3w_0z}$ at high-$z$. We, therefore, do not consider this parameterization in our analyses.} by assuming arbitrary values of $w_0$, $w_a$ and $T_0^{x} = 10^{-2}T_0^{\rm{CMB}}$, where  $T_0^{\rm{CMB}} = 2.73$ K. From this analysis, it is clear that an important point for the thermodynamic fate of the universe is to know how long the dark energy temperature will take to become the dominant temperature of the universe. A basic difficulty in estimating such a time interval, however, is that the present-day dark energy temperature has not been measured, being completely unknown.

\begin{figure}
\centerline{\psfig{figure=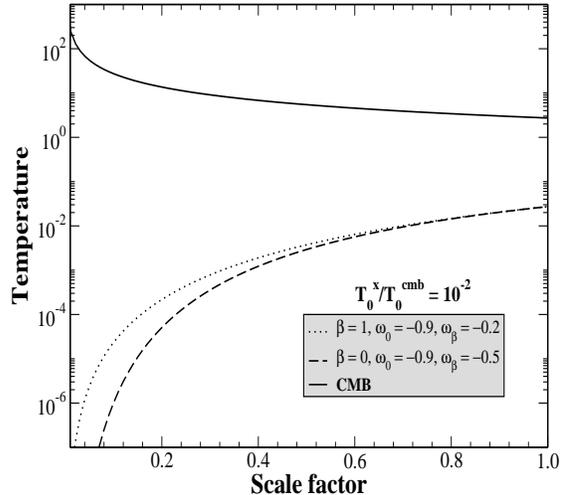,width=3.1truein,height=2.9truein,angle=-90}
\hskip 0.1in}
\caption{Temperatute evolution law for parameterizations P2 ($\beta = 0$) and P3 ($\beta = 1$) assuming some arbitrary values of $w_0$, $w_\beta$ and $T_0^{x} = 10^{-2}T_0^{\rm{CMB}}$, where  $T_0^{\rm{CMB}} = 2.73$ K. For comparison, we also show the CMB temperature curve (solid line).}
\end{figure}

By considering that the chemical potential for this $w(a)$-fluid is null (as occurs for $w = 1/3$), the Euler's relation defines its specific entropy, i.e.,
\begin{equation} \label{euler}
\sigma \equiv \frac{S_x}{N} = \frac{\rho_x + p_x}{nT}\;.
\end{equation}
Now, by combining the above equations, it is straightforward to show that the product $\rho_x a^3/T \equiv \rm{const.}$, so that
\begin{equation}\label{entropy}
S_x \propto (1 + w)\;.
\end{equation}
For a constant EoS parameter, the above expression recovers some of the results of Ref.~\cite{alcaniz}. Note also that the vacuum entropy is zero ($w = -1$) whereas for phantom dark energy ($w  < 1$), which violates all the energy conditions~\cite{phantom}, the entropy assumes negative values being, therefore, meaningless (For a discussion on the behavior of a phanton fluid with nonzero chemical potential, see~\cite{ademir}. See also \cite{gonzalez} for an alternative explanation in which the temperature of the phantom component takes negative values and \cite{thermo} for other thermodynamic analyses of dark energy).

Two cases of interest arise directly from Eq. (\ref{entropy}). The case in which $S_x = {\rm{const.}}$ implies necessarily that $w_\beta = 0$ for all the above parameterizations\footnote{Note that time-dependent EoS parameterizations without the constant term $w_0$ are imcompatible with the case $S_x = {\rm{constant}}$.}. The second case is more interesting, with the $w(a)$-fluid mimicking a fluid with bulk viscosity where the viscosity term is identified with the varying part of the dark energy pressure $\Pi$. For this latter case, we note that the positiveness of $S_x$ implies that
\begin{equation} \label{c1}
w_a \geq -\frac{1+w_0}{f(a)}\;,
\end{equation}
which clearly is not defined at $a = 1$, where $w = w_0$.
Finally, by combining Eqs.~(\ref{euler}) and (\ref{entropy}) with the conservation of the particle number density shown earlier, we obtain from the second law of thermodynamics that
\begin{equation} \label{c2}
S^{\alpha};_{\alpha} \propto \dot{w} \geq 0\;
\end{equation}
or, equivalently,
%\begin{equation}  \label{c2}
$w_a \leq 0$.
%\end{equation}

\begin{figure}[t]
\centerline{\psfig{figure=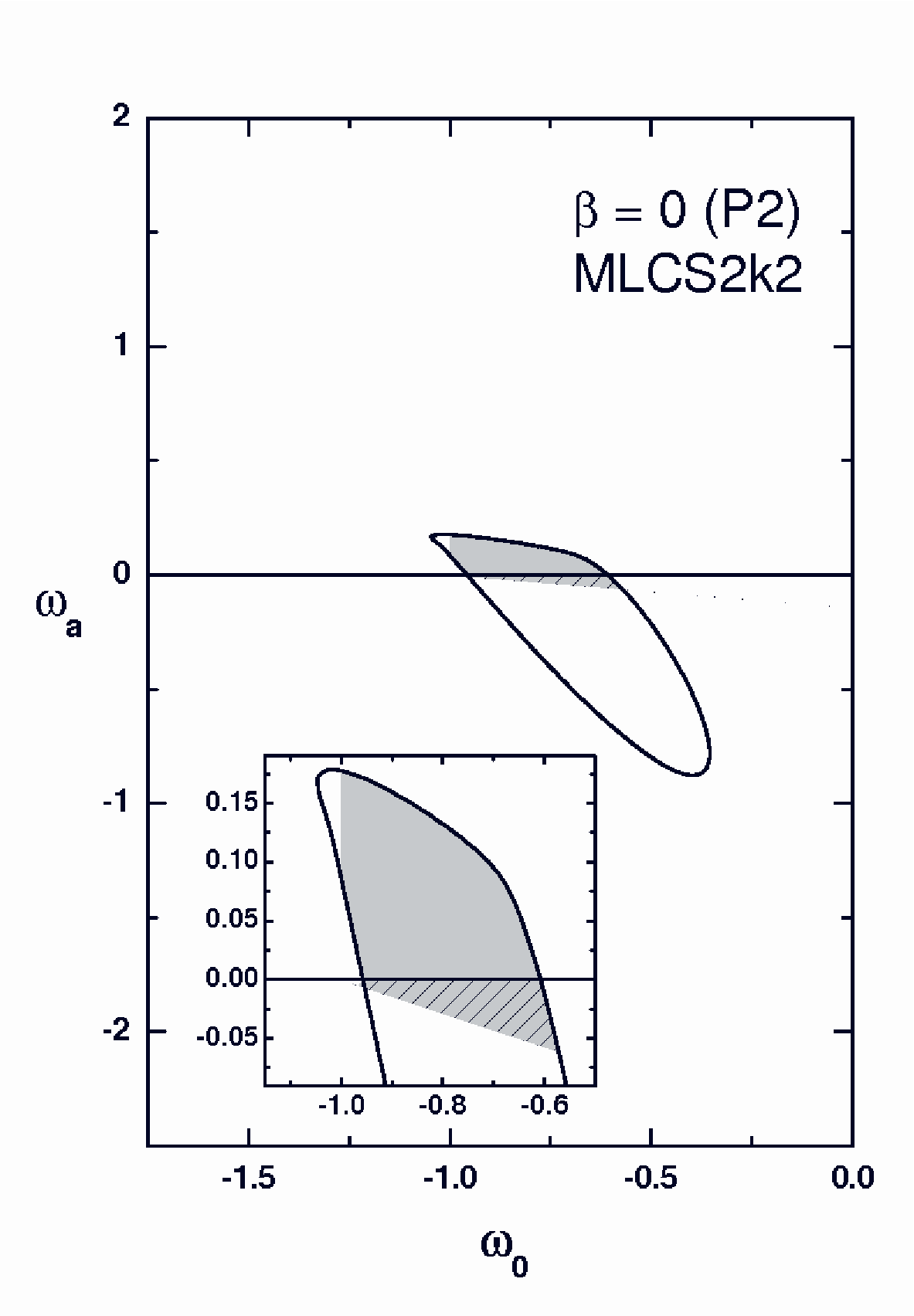,width=1.85truein,height=3.2truein,angle=0}
\psfig{figure=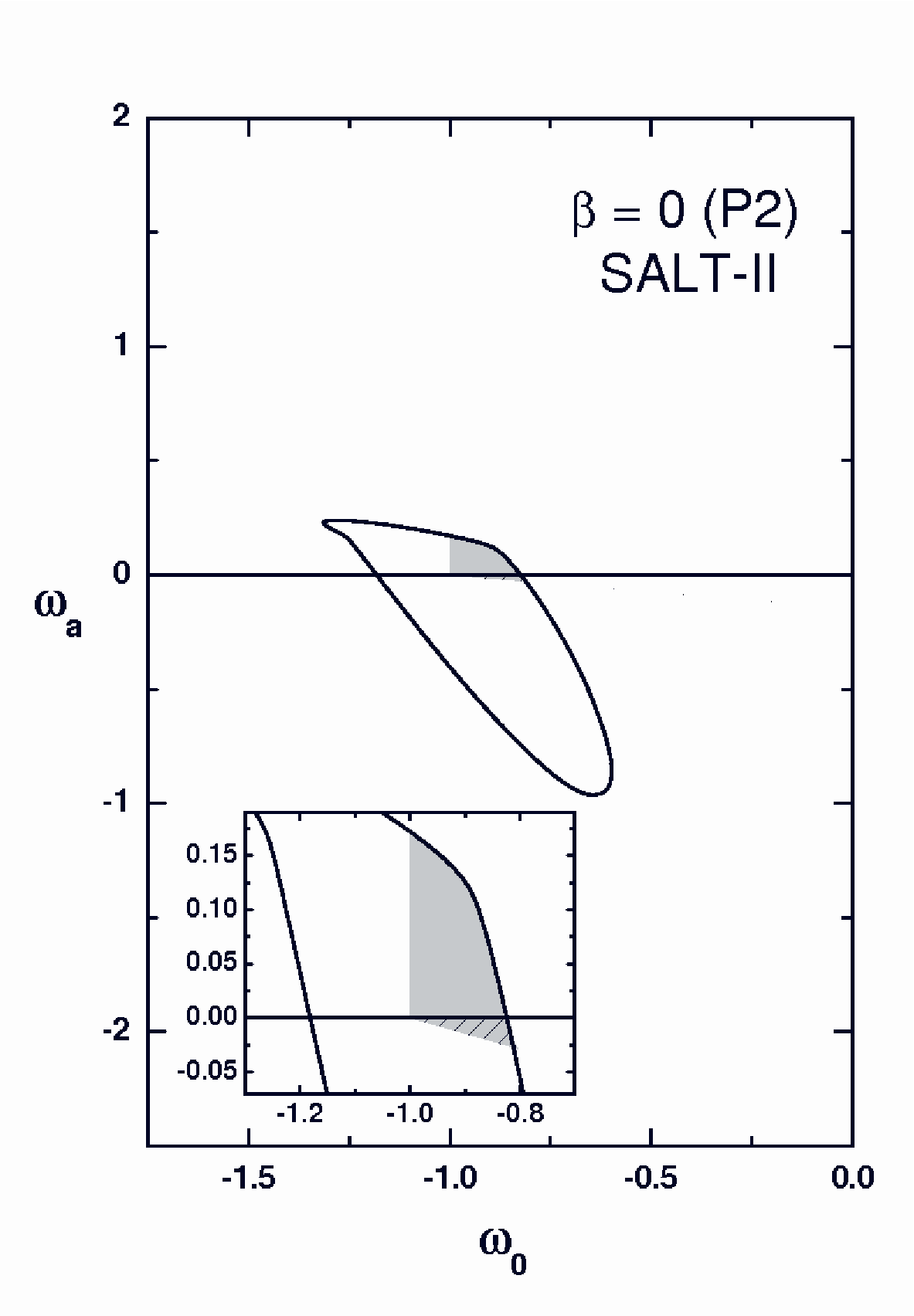,width=1.85truein,height=3.2truein,angle=0}}
%\hskip 0.1in}
\caption{Contours of $\Delta \chi^2 = 6.17$ in the parametric plane $w_0$ - $w_a$ for P2 $(\beta = 0)$. The light gray area represents the thermodynamic constraint of Eq. (\ref{c1}) whereas the small hachured area is the resulting parametric space when the constraint from the second law of thermodynamics [Eq. (\ref{c2})] is added to the analysis.}
\end{figure}

\section{Constraints on the $w_0 - w_\beta$ plane}

In this section, we combine the above physical constraints (\ref{c1}) and (\ref{c2}) with current observational data in order to impose bounds on the dark energy parameters. We  use one of the most recent SNe Ia data sets available, namely, the  SDSS compilation discussed in Ref.~\cite{sdss}. This compilation comprises 288 SNe Ia and uses both SALT2~\cite{salt2} and MLCS2k2~\cite{mlcs2k2} light-curve fitters~(see also \cite{friemanON} for a discussion on these light-curve fitters) and is distributed in redshift interval $0.02 \leq z \leq 1.55$. Along with the SNe Ia data, and  to help break the degeneracy between the dark energy parameters $w_0$ and $w_\beta$ we use the BAO~\cite{bao} and shift parameters~\cite{wmap}
 \begin{subequations}
\begin{equation}
{\cal{A}} = D_V{\sqrt{\Omega_{\rm{m}} H_0^2} \over {z_{\rm{BAO}}}}  = 0.469 \pm 0.017 \;,
\end{equation}
\begin{equation}
{\cal{R}} = \Omega_{\rm{m}}^{1/2}r(z_{\rm{CMB}}) = 1.71 \pm 0.019\;,
\end{equation}
\end{subequations}
where $D_V = [r^2(z_{\rm{BAO}}){z_{\rm{BAO}}}/{H(z_{\rm{BAO}})}]^{1/3}$ is the so-called dilation scale, defined  in terms of the dimensionless comoving distance $r$,  $z_{\rm{BAO}} = 0.35$ and $z_{\rm{CMB}} = 1089$. In our analyses, we minimize  the function $\chi^2 = \chi^{2}_{\rm{SNe}} + \chi^{2}_{\rm{BAO}} + \chi^{2}_{\rm{CMB}}$, which takes into account all the data sets mentioned above and marginalize over the present values of the matter density $\Omega_{\rm{m}}$ and  Hubble parameters $H_0$. % (see, e.g., \cite{analysis} for more details on this kind of statistical analysis).

\begin{figure}[t]
\centerline{\psfig{figure=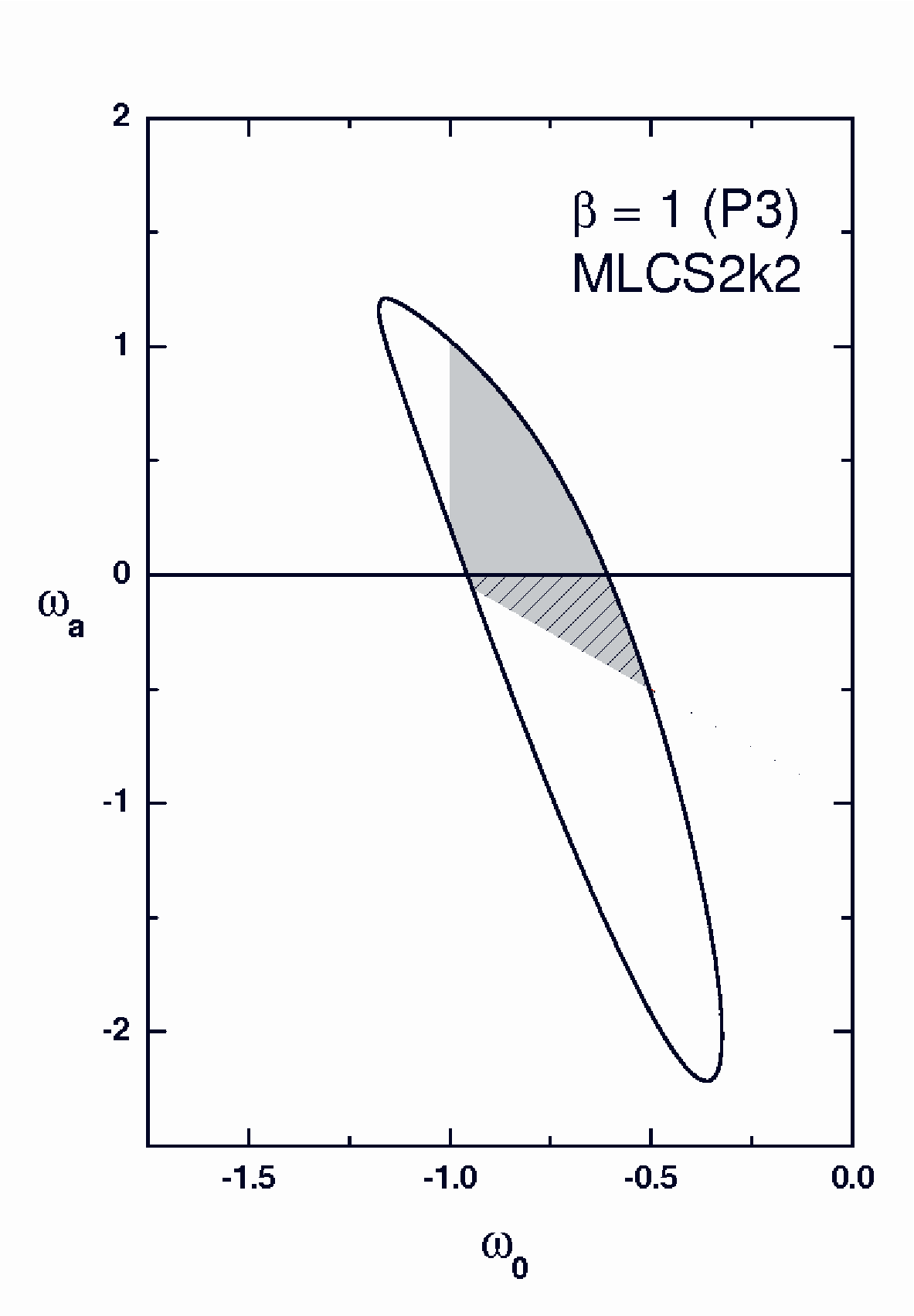,width=1.85truein,height=3.2truein,angle=0}
\psfig{figure=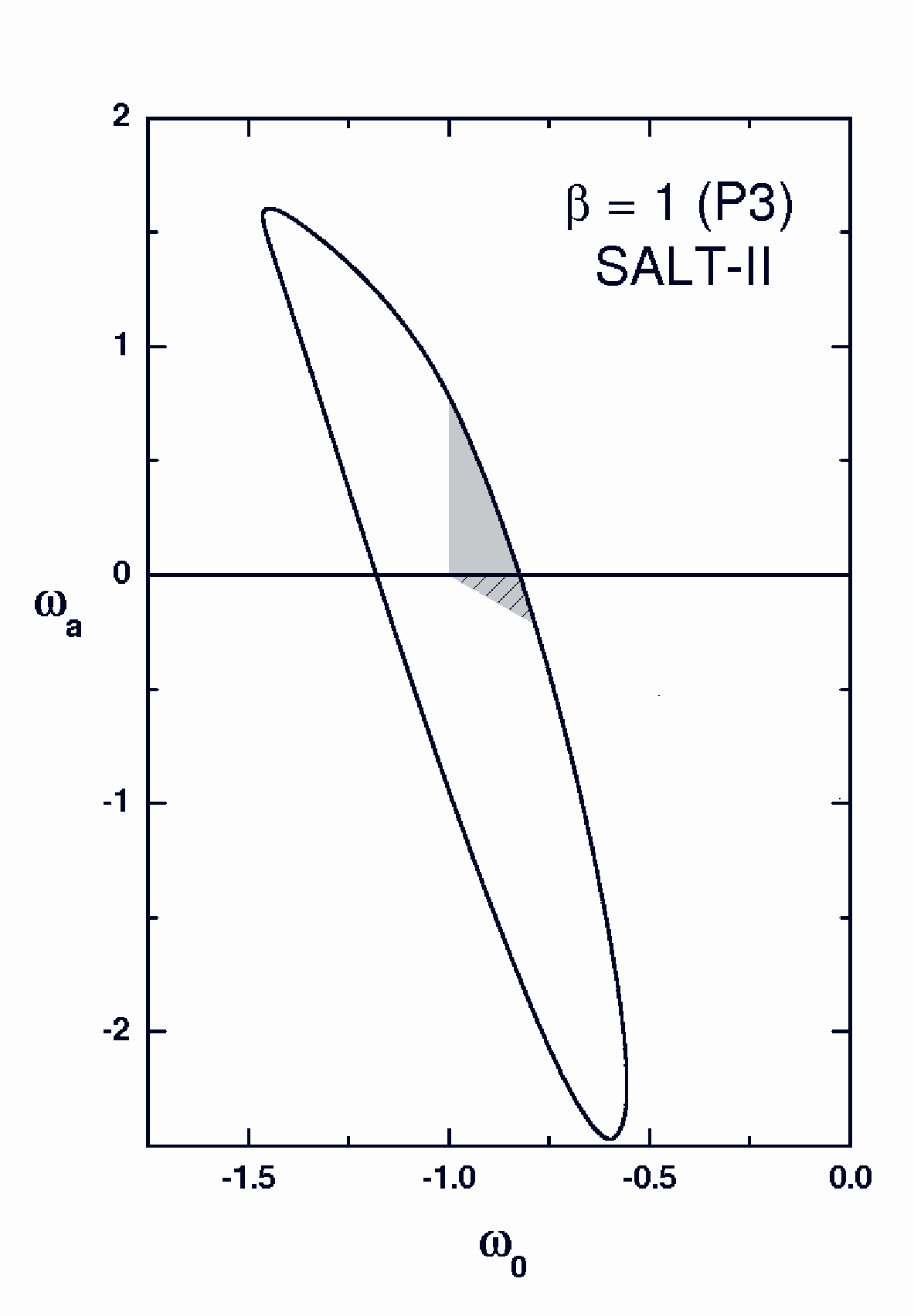,width=1.85truein,height=3.2truein,angle=0}}
%\hskip 0.1in}
\caption{The same as in Figure 2 for parameterization P3 ($\beta = 1$).}
\end{figure}

Figures 2 and 3 show the main results of our joint analyses. We plot contours of $\Delta \chi^2 = 6.17$ in the parametric space $w_0$ - $w_a$ for P2 and P3, respectively. The light gray region displayed in the plots stands for the physical constraint (\ref{c1}). Since this inequality is a function of time, the region is plotted by assuring its validity from $a = 10^{-4}$ up to today at $a = 1$. The resulting parametric space, when all the observational data discussed above are combined with the constraints (\ref{c1}) and (\ref{c2}), corresponds to the small hachured area right below the $w_a = 0$ line.  These results clearly illustrate the effect that the  thermodynamic bounds discussed in the previous section may have on the determination of the dark energy EoS parameters. In particular, we note that the resulting allowed regions are even tighter for the logarithmic parameterization P2 than for P3 (CPL). Since the SALT2 compilation allows for more negative values of $w_0$, the  joint constraints involving this SNe Ia sub-sample are also more restrictive (Figs. 2b and 3b).  For completeness, we display in Table I the changes in the 2$\sigma$ estimates of $w_0$ and $w_a$ due to the thermodynamic bounds (\ref{c1}) and (\ref{c2}).

\begin{table}[t]   \label{n-Table}
\begin{center}
%\begin{tabular}{|l|c|c|c|}
\begin{tabular}{lccc}
\hline \hline \\
Test& &$w_0$ & $w_a$\\
\hline \hline \\
SNe Ia (MLCS2k2)\footnote{+ CMB + BAO}....................& P2 & $-0.78^{+0.33}_{-0.21}$ & $0.02^{+0.14}_{-0.66}$\\
SNe Ia (MLCS2k2)$^a$ + T\footnote{Thermodynamic constraints [Eqs. (11) and (12)]}..........& P2 &$ -0.77^{+0.16}_{-0.14}$& $0.00^{+0.00}_{-0.05}$\\
SNe Ia (SALT2)$^a$........................& P2 & $-1.05^{+0.34}_{-0.20}$& $0.13^{+0.09}_{-0.80}$\\
SNe Ia (SALT2)$^a$ + T$^b$...............& P2 & $-0.99^{+0.14}_{-0.01}$& $0.00^{+0.00}_{-0.02}$ \\
SNe Ia (MLCS2k2)$^a$....................& P3 & $-0.81^{+0.37}_{-0.29}$ & $0.18^{+0.99}_{-1.81}$\\
SNe Ia (MLCS2k2)$^a$ +   T$^b$..........& P3 & $-0.77^{+0.22}_{-0.14}$& $0.00^{+0.00}_{-0.45}$\\
SNe Ia (SALT2)$^a$........................& P3 & $-1.09^{+0.41}_{-0.30}$& $0.52^{+0.96}_{-2.26}$ \\
SNe Ia (SALT2)$^a$ + T$^b$...............& P3 & $-0.99^{+0.17}_{-0.01}$& $0.00^{+0.00}_{-0.18}$\\
\hline \hline
\end{tabular}
\end{center}
%\caption{Change in the best-fit values of $w_0$ and $w_a$ due to the thermodynamic bounds (\ref{c1}) and (\ref{c2}).}
\end{table}

\section{Conclusions}

In spite of its fundamental importance for an actual understanding of the evolution of the universe, the relevant physical properties of the dominant dark energy component remain completely unknown. In this paper we have investigated some thermodynamic aspects of this energy component assuming that its constituents are massless quanta with a general time-dependent EoS parameter $w(a)$. We have discussed its temperature evolution law and derived constraints from the second law of thermodynamics on the values of $w_0$ and $w_a$ for a family of $w(a)$ parameterizations given by Eq. (\ref{pb}). When combined with current data from SNe Ia, BAO and CMB observations, we have shown that such constraints provide very restrictive limits on the parametric space $w_0$ - $w_a$ (see Figs. 2 and 3).

Finally, it is also worth mentioning that in the present analysis we have assumed that the chemical potential $\mu$ for the $w(a)$-fluid representing the dark energy is null. A more general analysis relaxing this condition ($\mu \neq 0$) is currently under preparation and will appear in a forthcoming communication.

\begin{acknowledgments}
The authors are very grateful to J. A. S. Lima for helpful discussions and a critical reading of the manuscript and to CNPq and CAPES for the grants under which this work was carried out.
\end{acknowledgments}


\begin{thebibliography}{30}

\bibitem{review} B. Ratra \& M.\ S.\ Vogeley, PASP  {\bf{120}}, 235 (2008); R. R. Caldwell \& M. Kamionkowski, Ann.\ Rev.\ Nucl.\ Part.\ Sci. {\bf{59}}, 397 (2009);  A. Silvestri \& M. Trodden, Rept.\ Prog.\ Phys. {\bf{72}}, 09690 (2009); M. Sami, Curr.\ Sci.\ {\bf{97}}, 887 (2009).

\bibitem{geral} E.~M.~Barboza, J.~S.~Alcaniz, Z.~H.~Zhu and R.~Silva, Phys.\ Rev.\  D {\bf 80}, 043521 (2009).

\bibitem{8} A. R. Cooray and D. Huterer, Astrophys. J. {\bf 513}, L95 (1999); P. Astier, Phys. Lett. B, {\bf 500}, 8 (2001); J. Weller and A. Albrecht, Phys. Rev D {\bf 65}, 103512 (2002).

\bibitem{14} G. Efstathiou, Mon. Not. Roy. Astron. Soc., {\bf 310}, 842 (1999).

\bibitem{15} M. Chevallier and D. Polarski, Int. J. Mod. Phys. D {\bf 10}, 213 (2001); E. V. Linder, Phys. Rev. Lett. {\bf 90}, 091301 (2003).

\bibitem{para} Y. Wang and P. M. Garnavich, Astrophys. J. {\bf 552}, 445 (2001); C. R. Watson and R. J. Scherrer, Phys. Rev. D {\bf 68}, 123524 (2003); P.S. Corasaniti et al., Phys. Rev. D {\bf 70}, 083006 (2004); V. B. Johri, astro-ph/0409161; Y. Wang and M. Tegmark, Phys. Rev Lett. {\bf 92}, 241302 (2004);   H. K. Jassal, J. S. Bagla, and T. Padmanabhan, Mon. Not. Roy. Astron. Soc. {\bf 356}, L11 (2005); E. M. Barboza Jr. and J. S. Alcaniz, Phys. Lett. B {\bf 666}, 415 (2008).

\bibitem{sdss} R. Kessler {\it et al.}, Astrophys. J. Suppl. Ser. {\bf{185}}, 32 (2009).

\bibitem{salt2} J. Guy {\it et al.}, Astron. Astrophys. {\bf{466}}, 11 (2007).

\bibitem{mlcs2k2} M. M. Phillips, Astrophys. J. {\bf{413}}, L105 (1993); A. G. Riess, W. H. Press, and R. P. Kirshner, Astrophys. J. {\bf{438}}, L17 (1995); S. Jha, A. G. Riess, and R. P. Kirshner, Astrophys. J. {\bf{659}}, 122 (2007).

\bibitem{bao} D. J. Eisenstein {\it et al.}, Astrophys. J. {\bf{633}}, 560 (2005).

\bibitem{wmap} D. N. Spergel et al. Astrophys. J. Suppl. Ser. {\bf{170}}, 377 (2007).

\bibitem{landau} D. Landau and E. M. Lifshitz, Fluid Mechanics (Pergamon Press, New York, 1959).

\bibitem{weinberg} S. Weinberg, Astrop. J. {\bf 168}, 175 (1971)

\bibitem{silva} J. A. S. Lima, A. S. M. Germano, Phys. Lett. A {\bf 170}, 373 (1992); R. Silva, J. A. S. Lima and M. O. Calv\~ao, Gen. Rel. Grav. {\bf 34}, 865 (2002).

\bibitem{alcaniz} J. A. S. Lima and J. S. Alcaniz, Phys. Lett. B {\bf 600}, 191 (2004).

\bibitem{janilo} J.A.S. Lima and J. Santos, Int. J. Theor. Phys. {\bf{34}}, 143 (1995).

\bibitem{phantom}R. R. Caldwell, Phys. Lett. B {\bf{545}}, 23 (2002); S. M. Carroll, M. Hoffman and M. Trodden, Phys. Rev. D {\bf{68}}, 023509 (2003); J.~S.~Alcaniz,  Phys.\ Rev.\  D {\bf 69}, 083521 (2004).

\bibitem{ademir}   J.~A.~S.~Lima  and S.~H.~Pereira, Phys. Rev. D 78, 083504 (2008);  S.~H.~Pereira and J.~A.~S.~Lima,  Phys.\ Lett.\  B {\bf 669}, 266 (2008).

\bibitem{gonzalez}  P.~F.~Gonzalez-Diaz and C.~L.~Siguenza,   Nucl.\ Phys.\  B {\bf 697}, 363 (2004).

\bibitem{thermo}   G.~Izquierdo and D.~Pavon,  Phys.\ Lett.\  B {\bf 633}, 420 (2006); N.~Bilic,  Fortsch.\ Phys.\  {\bf 56}, 363 (2008); Y.~S.~Myung,  Phys.\ Lett.\  B {\bf 671}, 216 (2009); E.~N.~Saridakis, P.~F.~Gonzalez-Diaz and C.~L.~Siguenza,  Class.\ Quant.\ Grav.\  {\bf 26}, 165003 (2009).

\bibitem{friemanON} J.~A.~Frieman, AIP Conf.\ Proc.\  {\bf 1057}, 87 (2008). arXiv:0904.1832.

\end{thebibliography}
\end{document}